\documentstyle[citesort,12pt]{article}
\def \bfgr #1{ \mbox {{\boldmath $#1$}}}
\begin{document}

\begin{titlepage}
\title{ {\Large \bf Inquiry for the $\Large ({\bfgr \pi}^{+}-{\bfgr \pi}^{-})$
Bound state Conversion in two $\Large {\bfgr \pi}^{0}$ as being due to the
Weinberg $\Large \bfgr \pi - \bfgr \pi$-interaction Lagrangian.}}
\author{{\large \bf G. G. Bunatian}\\
{\it Joint Institute for Nuclear Research, 141980, Dubna, Russia.}\\
\vspace{1.5cm}
\\
Gevorg G. Bunatian\\
Joint Institute for Nuclear Research (JINR), \\
141980, Dubna,
Moscow Reg., Russia.\\
E-mail:  BUNAT@CV.JINR.DUBNA.SU\\
FAX:  (7-096)-21-65084\\
FAX:  (7-095)-975-23-81 \\
Phon:  (095) 924-39-14 \\
Telex:  911621 DUBNA SU}
\maketitle
\end{titlepage}
%\newpage

\begin{center}
{\bf Abstract}
\end{center}

In the work presented, the decay of the pionium, that is the
$\large (\bfgr \pi ^{+}\bfgr \pi ^{-})$ bound state, into two
$\large \bfgr \pi ^{0}$ is studied, the
$\large \bfgr \pi\bfgr \pi$-interaction causing this transition being
described by the underlying Weinberg lagrangian. The calculation with such a
 $\large \bfgr \pi\bfgr \pi$-lagrangian being carried out, the $\large
 \bfgr \pi$-meson size $\large r_{0}$ emerges to be allowed for and
 occurs in the final result. The bound
 $\large (\bfgr \pi ^{+}\bfgr \pi ^{-})$-system itself is presumed to be
 due to the instantaneous Coulomb interaction and is treated consistently
 nonrelativistic, the Bethe-Solpeter equation being utilized. Along
 calculating, the terms of the lowest order in fine structure constant
 $\large \bfgr \alpha$ and the terms $\large \sim ln(r_{0})$ are
  retained. The obtained pionium lifetime $\large \bfgr \tau$ is thought to
 be compatible with the conceivable future experimental data.  The dependence
of the results on the effective lagrangian parameters is visualized.
The investigation carried out persuades us that just the whole form of the
genuine $\large \bfgr \pi\bfgr \pi$-interaction determines the pionium
lifetime, but not simply the  $\large \bfgr \pi\bfgr \pi$ scattering
lengths. The inquiry into pionium decaying promotes to specify the
 validity of various $\large \bfgr
 \pi\bfgr \pi$-interaction descriptions.\\
\\
\\


\\
\\

\newpage

\begin{center}
{\large \bf 1. Introduction. Agenda of $\large \bfgr \pi \bfgr
\pi$-interaction.}
\end{center}

At present, the stringent knowledge of $\pi \pi$-interaction is well
understood to be of fundamental value in its own right as well as for the
reliable treatment of the various phenomena, where pionic degrees of freedom
prove to be substantial. Pion being the lightest and, properly speaking,
simplest among strong-interacting particles, an inquiry into pion-pion
interaction spread the way to visualization of the main features of hadrons
interactions in general, in their immense complexity \cite{w1,w2,g1,lg,nj}.
At the same time, the pion-pion interaction is bound to be allowed for in
describing the hot and dense hadronic systems abundant in pions which are
known to be produced in heavy ions colliding \cite{c1} at high enough
incident energy, the baryon number being rather negligible when
compared with the number of genuine mesons. Even so, in treating the nuclear
matter at large density and temperature, the phenomena non-linear in meson
fields, that is the meson-meson interactions, are realized to play the
crucial role, especially when the feasible phase transitions caused by the
mesonic degrees of freedom softening are investigated \cite{c2}. Thus, to
repose full confidence in adequacy of our perception of such systems
behaviour, the pion-pion interaction must be properly accounted for, in
particular along calculating the respective thermodynamic characteristics.
Thereby, in all the cases, we must certainly conceive the pion-pion
interaction to be provided by the well specified trustworthy lagrangian, but
not in the least simply just by pion-pion scattering lengths.

Nowadays, in the lack of the pions interactions description strictly worked
out from the first principles, we are in possession of the pion-pion
interaction lagrangians \cite{w1,w2,g1,lg,nj} which are thought to be as good
as effective, obtained in frameworks of some plausible models, $QCD$-motivated
at best. Consequently, there is to appeal to the experimental investigations
the reliable information about the $\pi \pi$-interaction can be disentangled
from.  Then, confronting the results of processing experimental data and of
theoretical calculations, we can check up the validity of a certain $\pi
\pi$-interaction description and subsequently ameliorate the latter.

Up to now, the trustworthy cognizance concerning $\pi \pi$-interaction has
been acquired, strictly speaking, solely from the analysis of the data
obtained in the $\pi N\rightarrow\pi \pi N$ reaction which was studied at
first time as far back as in 1965 \cite{bbo} near threshold
(${\varepsilon}_{\pi}\sim 200 - 300 \;
MeV$) and afterwards for manifold
incident pion energies, up to ${\varepsilon}_{\pi}\sim 1 - 2 \;
GeV$ as well
(see, for instance, \cite{bb2,bb3}). The results of profound processing these
 experimental data carried out in the series of investigations
 \cite{bb2,bb3,bb1} make us visualize the effective lagrangians asserted in
 \cite{w1,w2,g1} are thought to be expedient to describe the $\pi
 \pi$-interaction, at least for low and middle pion energy,
${\varepsilon}_{\pi}\sim m_{\pi}$. Unfortunately, the inescapable involvement
of strong pion-nucleon interactions in such a process put a bound to an
attainable reliability of the pure $\pi \pi$-interaction description because,
on one hand, it is as good as impossible to get rid of the strong $\pi
N$-interaction effect in the experimental measurements, and, on the other
hand, one will scarcely maintain that a theoretical calculation can refine
unambiguously the $\pi \pi$-interaction from $\pi N$ one in the reaction $\pi
N\rightarrow\pi \pi N$ treatment. Thus, the further development of $\pi
\pi$-interaction description by means of the profound effective lagrangian
\cite{lg,nj}, or my be along other approaches (see, for instance,
\cite{bb4}), calls on new experiment. For that matter, at first thought, the
$K_{e4}$-decay, $K\rightarrow\nu e \pi \pi$, \cite{kls} might appear of
being fruitful to learn directly  the pure $\pi \pi$-interaction occurring in
the final state, but it has to realize the weak-decay vertex itself is not
concisely known and needs to be approved in its own right \cite{com}. Thus, as
yet, the reaction $\pi N\rightarrow\pi \pi N$ was and has been, as a matter
of fact, the unique source of the data to check our perception about the $\pi
\pi$-interaction.

In the light of the aforesaid agenda, the advent of the
experiments dealing with the pure $\pi \pi$-interaction, without imposition
of other strong (or weak) interactions, proves to be extremely desirable.

\begin{center}
{\large \bf 2. Up to now Pionium Treatment.}
\end{center}

Long since, the inquiry into the properties of the ${\pi}^{+}{\pi}^{-}$ bound
state, pionium, had been understood of being very instructive to study of the
pure $\pi \pi$-interaction, free of effect of any other strong or weak
interactions \cite{n1}. The feasible measurement of the pionium lifetime
having been firstly considered in the early investigations \cite{n1}, setting
up the correspondent experiment has been profoundly elaborated in Refs.
\cite{n2,n3}, and the respective investigations are for now already under way
\cite{n3}, the results are liable to arrive in the nearest future.

Pionium typifies the bound hadrons systems their very coming into existence
is due to electromagnetic interactions, but their decay is, as a matter of
fact, caused by strong interactions. All the time ago, as far back as  in
1954, the handy semiquantitative approach to treat such systems was set
out \cite{dg}, the strong interaction corrections to energy levels and wave
functions of the $\pi$-atom, the ${\pi}^{-}P$ bound state, as well as the
transition rate ${\pi}^{-}P\rightarrow {\pi}^{0}N$ being expressed
 through the free pion-nucleon scattering lengths $a^T_L$ and the
$\pi$-atom wave function at the origin ${\psi}(0)$. Here, $T,L$ indices
denote various isotopic and angular states. Subsequently following this
method, the pionium lifetime, that is the
${\pi}^{+}{\pi}^{-}\rightarrow{\pi}^{0}{\pi}^{0}$ reaction rate, in
 the ground state was asserted in Refs. \cite{n1} to be the simple
plain function
\begin{equation}
{\tau}^{-1}=\frac{16\pi}{9}\sqrt{\frac{2 {\Delta}m}{m}}|a^0_0-a^2_0|^2\cdot
|{\psi}(0)|^2
\label{t}
\end{equation}
of the $s$-wave $\pi \pi$-scattering lengths $a^0_0 , a^2_0$, the pionium
wave function at the origin ${\psi}(0)$, and the mass difference
${\Delta}m=m-m_{0}$, $m$ being charge pion mass. Thus, if the original
approach of Ref. \cite{dg} had been strictly valid in the pionium case, all
we need for the pionium lifetime precise calculation would have been the
exact values of the quantities $|a^0_0-a^2_0|, \; |{\psi}(0)|$,
 and ${\Delta}m$.  It is to take cognizance of the fact that only the
difference of the scattering lengths would have come into picture, regardless
the complete form of the genuine $\pi \pi$-interaction. This is due to the
main original presumption of the approach of Ref. \cite{dg} that irrespective
to the $\pi \pi$-interaction form the calculation of probability of the
pionium decay into two ${\pi}^{0}$ is quite equivalent to the calculation of
the annihilation probability of an free pair  ${\pi}^{+}{\pi}^{-}$ with zero
momenta into two ${\pi}^{0}$,
${\pi}^{+}{\pi}^{-}\rightarrow{\pi}^{0}{\pi}^{0}$, the initial state density
 being not the free particles state density, but the density of state of
the particles in the bound state of pionium $|{\psi}(0)|^2$. Up to now, the
authors of all the succeeding investigations
\cite{dg1,dg2,wy,ras,eul,bpt,lub,sil} have been taking for granted that the
pionium lifetime formula (\ref{t}) asserted according to \cite{dg} in Ref.
\cite{n1} holds true strictly, and all the efforts were devoted to
acquire somehow the precise values of the quantities $a^T_L , {\psi}(0)$, the
pure point-like Coulomb nonrelativistic ${\psi}(0)$ value and the
free-particles scattering lengths $a^T_L$ values gained according to Refs.
\cite{w1,w2,g1,lg} being assumed as a starting point in all the calculations.
Then, there was to calculate the corrections to that ${\psi}(0)$ value,
especially due to strong interactions, and simultaneously the
$a^T_L$-modifications on account of strong and electromagnetic interactions
in the coupled ${\pi}^{0}{\pi}^{0}, {\pi}^{+}{\pi}^{-}$ channels.

In the several
investigations \cite{dg1,dg2,wy,ras,eul,bpt} the various effective
potentials were managed to describe this strong $\pi
\pi$-interaction. The most profound calculations within such a potential
approach have been carried out in Ref. \cite{wy} and especially in the work
\cite{ras}, where the aforesaid corrections have been thoroughly calculated
in the framework of the model of the two-channels
${\pi}^{0}{\pi}^{0}, {\pi}^{+}{\pi}^{-}$ system, the effective range
approximation being used to account for the strong pion-pion interaction.
Thereby, once an effective radius is chosen (equal in both channels), the
strong potentials in the channels are determined merely just by the
correspondent scattering lengths $a^T_0$. In such a calculation, the
electromagnetic corrections are due to the pion mass difference
in different channels along with the Coulomb interaction imposition in the
${\pi}^{+}{\pi}^{-}$ channel.  The coupled Schr\"odinger  equations
determining the pions wave functions in the coupled channels having been
solved, the corrected, generalized scattering lengths as well as
subsequently corrected ${\psi}(0)$ values are obtained which must be
substituted in the original formula (\ref{t}) for $\tau$ to acquire
its eventual corrected value. The scrutinized corrections to $a^{T}_{L}$
values (and to ${\psi}(0)$ sa well) proved to amount not more than few
per-cents, being substantially less than the
uncertainties in the $a^{T}_{L}$ predictions following from Ref. \cite{lg}, as the
authors of \cite{ras} have inferred.

Unlikely the effective potential approach of the Refs.
\cite{dg1,dg2,wy,ras,eul,bpt}, the investigation \cite{lub} utilized the
Bethe-Solpeter equation to allow for the effect of strong interactions on
the ${\psi}(0)$ value in the pionium lifetime (\ref{t}) (via the pionium
eigenstate energy shift ${\Delta}E$), the correction proving to be rather
negligible.

The $\tau$ (\ref{t}) value modification on account of pionium
relativistic treatment, especially the allowance for retardation effect in the
${\pi}^{+}{\pi}^{-}$ electromagnetic interaction has been found $\sim 1\%$
in Ref. \cite{sil}. Thereby, the scattering lengths difference $a_0^0-a_0^2$
was presumed of rendering the total strong interaction responsible of
${\pi}^{+}{\pi}^{-}\rightarrow{\pi}^{0}{\pi}^{0}$ transition, likewise
in all the aforecited investigations \cite{dg1,dg2,wy,ras,eul,bpt},
in spite of treating the retardation effect in the ${\pi}^{+}{\pi}^{-}$
system which implies the  ${\pi}^{+}{\pi}^{-}$ relative velocity to
be comparable with light velocity $c$.

Profound as are all the before discussed calculations of the quantities
$a^T_0 , {\psi}(0)$, we ought to realize the expression (\ref{t}) itself,
insofar as originating from the very plausible, but semiquantitative approach
\cite{dg}, is, properly speaking, as good as semiquantitative  in turn. But
this did not mean to say that any results obtained accordingly to the
method set out in Ref. \cite{dg} must be regarded as untenable and
scarcely able to describe experimental data with high enough accuracy.
There is to visualize  the validity and accuracy of this very
approach in each certain treated case is caused crucially by the form of
the genuine strong interaction inducing the bound hadronic system decay.
The very germ of the idea set forth in Ref. \cite{dg} makes us comprehend
the approach of \cite{dg} itself will hold true with high precision, if the
hadron-hadron interaction is as good as point-like and constant, especially
momentum-independent which is thought to be well acceptable for
$P{\pi}^{-}$-interaction in the $s$-state in \cite{dg}, but not in the least
for the $\pi \pi$-interactions asserted and used in Refs.
\cite{w1,w2,g1,lg,bb2,bb3,bb1,bb4}. Consequently, pionium properties being
studied, we must refrain from to pursue the way paved in Ref. \cite{dg}
and reject, in turn, the handy expression (\ref{t}) for pionium lifetime.

\begin{center}
{\large \bf 3. Interactions inducing pionium decay into two $\large \bfgr
\pi ^{0}$.}
\end{center}

According to our lights, a general aim of the theoretical investigations of
 pionium lifetime is to visualize whether a certain form of the $\pi
 \pi$-interaction is eligible to provide the experimental $\tau$ value. In
 the work presented, we set out the calculation of $\tau$, the $\pi
 \pi$-interaction being determined by the Weinberg lagrangian according to
 Refs. \cite{w1,w2,g1}. The probability of two-photons pionium annihilation,
 ${\pi}^{+}{\pi}^{-}\rightarrow 2\gamma$, being practically negligible when
 compared with decay probability due to strong interaction, will not be
  discussed henceforth.

We treat pionium as the beforehand prepared ${\pi}^{+}{\pi}^{-}$ bound state
which is stable when strong pion fields interaction is turned off. The
coupling of this state, pionium field, to the charge (complex) pion field is
implemented via the virtual decay of the ${\pi}^{+}{\pi}^{-}$ bound state
$|{\cal D}_{\lambda}>$, pionium or di-meson, into free ${\pi}^{+}{\pi}^{-}$ pair :
\begin{equation}
{\pi}^{+}+{\pi}^{-}\leftarrow |{\cal D}_{\lambda}>
\label{d}
\end{equation}
In our nowaday consistently nonrelativistic approach, we presume the
formation of the initial ${\pi}^{+}{\pi}^{-}$ bound state ${|\cal
D}_{\lambda}>$ is caused by pure-instantaneous potential interaction $U({\bf
y}_1 , {\bf y}_2)$, where ${\bf y}_1 , {\bf y}_2$ are spatial coordinates of
the ${\pi}^{+}({\bf y}_1 , t), {\pi}^{-}({\bf y}_2 , t)$ mesons composing the
pionium, the time coordinates coinciding. Subsequently, the vertex operator
\begin{eqnarray}
\hat{\cal L}_{\cal D}= - [{\pi}^{+}({\bf y}_1 , t) {\pi}^{-}({\bf y}_2 , t) +
{\pi}^{-}({\bf y}_1 , t) {\pi}^{+}({\bf y}_2 , t)] \hat{\cal F}({\bf y}_1 ,
{\bf y}_2 , t)  , \label{5o}\\
\hat{\cal F}({\bf y}_1 , {\bf y}_2 , t) = \sum_{\lambda}[c_{\lambda}
{\cal F}_{\lambda}({\bf y}_1 , {\bf y}_2 , t) +
 c_{\lambda}^{+}{\cal F}_{\lambda}^{*}({\bf y}_1 , {\bf y}_2 , t)]
 \label{5}
\end{eqnarray}
\setlength{\unitlength}{1cm}
\begin{picture}(16.,3.)
\thicklines
\put(8,1.5){\circle{1.}}
\put(7.85,1.35){$\hat {\bf\cal F}$}
\thinlines
\multiput(8.5,1.43)(0,0.12){2}{\line(1,0){0.7}}
\put(9.34,1.35){${\bf {|\cal D}_{\lambda}>}$}
\thicklines
\put(7.6,1.7){\vector(-3,2){1.2}}
\put(7.6,1.3){\vector(-3,-2){1.2}}
\put(6.,2.55){${\bfgr \pi}^{\bf +}$}
\put(5.92,0.45){${\bfgr \pi}^{\bf -}$}
\end{picture}
\\
renders the virtual pionium state $|{\cal D}_{\lambda}>$ decay into a free
${\pi}^{+}{\pi}^{-}$ pair. Here, ${\pi}^{\pm}({\bf y} ,t)$ are charge pion
field operators, whereas $\hat{\cal F}({\bf y}_1 , {\bf y}_2 , t)$ stands
for the pionium field, the quantities $c_{\lambda}, c_{\lambda}^{+}$ being the
pionium  production and distraction operators in the state $\lambda$.
So far as the interaction $U({\bf y}_1 , {\bf y}_2)$ is instantaneous, all
the fields operators in (\ref{5o}, \ref{5}) act at the same time point $t$.
In our calculation, the common relations are adopted
\begin{eqnarray}
{\pi}^{+}(x) = \frac{1}{\sqrt{2}}(\pi _1(x) + i\pi _2(x)), \; \; {\pi}^{-}(x)
 = -({\pi}^{+}(x))^{*}, \; \; {\pi}^{0}(x) = {\pi}_3(x), \nonumber \\
{\pi}^{+}(x) = \sum_{{\bf p}}\frac{1}{\sqrt{2{\varepsilon}_{{\bf p}}}}
[a_{{\bf p}} e^{-it{\varepsilon}_{{\bf p}}+{\bf p}{\bf x}} +
b^{+}{{\bf p}} e^{it{\varepsilon}_{{\bf p}}-{\bf p}{\bf x}}],
\label{1ab}
\end{eqnarray}
the operator $a_{{\bf p}}$ destructing of ${\pi}^{+}$-meson and $b^{+}_{{\bf
p}}$ producing ${\pi}^{-}$-meson. The vertex functions
${\cal F}_{\lambda}({\bf y}_1 , {\bf y}_2 , t)$ in (\ref{5}) and the
correspondent pionium eigenenergies $E_{\lambda}$ in the states
$\lambda$ are well known (see, for instance, Refs. \cite{ll,sw}) to be
determined by the homogeneous Bethe-Solpeter equation
\begin{equation}
{\cal F}_{\lambda}({\bf y}_1 , {\bf y}_2 , t) = U({\bf y}_1 , {\bf y}_2)
 \cdot \int dt' \int d{\bf y}'_1 \int d{\bf y}'_2 {\it D}(y_1-y'_1)
 {\it D}(y_2-y'_2) {\cal F}_{\lambda}({\bf y}'_1 , {\bf y}'_2 , t'),
\label{5a}
\end{equation}
where
\begin{equation}
{\it D}(x) = \frac{1}{i(2\pi)^4}\int \frac{d^{4}k \cdot e^{ikx}}
{k^2-m^2+i\delta}
\label{5aa}
\end{equation}
is the usual pion propagator. In presumed non-relativistic approach, the
vertex function ${\cal F}_{\lambda}({\bf y}_1 , {\bf y}_2 , t)$ proves to
be reduced as follows (see, for instance, Refs. \cite{ll,sw} and also
\cite{ab})
\begin{equation}
{\cal F}_{\lambda}({\bf y}_1 , {\bf y}_2 , t) = - i{\cal N} \cdot
 U({\bf y}_1 , {\bf y}_2) \cdot {\Phi}_{\lambda}({\bf y}_1 , {\bf y}_2 , t) ,
\label{5b}
\end{equation}
where ${\Phi}_{\lambda}({\bf y}_1 , {\bf y}_2 , t)$ is the
non-relativistic ${\pi}^{+}{\pi}^{-}$ system wave function. The function
${\cal F}_{\lambda}$ being determined by the homogeneous equation (\ref{5a}),
the normalization factor ${\cal N}$ emerges in (\ref{5b}) which calculation
we defer for a while. The wave function ${\Phi}_{\lambda}({\bf y}_1 , {\bf
y}_2 , t)$ of such a nonrelativistic system is known (see, for instance,
\cite{shl}) to be the product \begin{eqnarray} {\Phi}_{\lambda}({\bf y}_1 ,
{\bf y}_2 , t) = {\psi}_{nl}({\bf z}) \cdot {\Psi}_{{\bf p}}({\bf R}) \cdot
e^{itE_{\lambda}}, \; \; E_{\lambda}=2m+ \frac{{\bf
P}^2}{4m}+{\varepsilon}_{nl} , \; {\lambda}=(nl,{\bf P}) \label{2}
\end{eqnarray}
of the depending on the center of mass coordinate ${\bf R}=({\bf y}_1+{\bf
y}_2)/2$ wave function
\begin{equation}
{\Psi}_{{\bf P}}({\bf R}) = \frac{1}{\sqrt{2E_{\lambda}}}e^{i{\bf RP}}
\label{3}
\end{equation}
of the free motion of the two-pion system as a whole with the total momentum
${\bf P}$, and the intrinsic pionium wave function ${\psi}_{nl}({\bf z})$
depending on the relative ${\pi}^{+}{\pi}^{-}$
coordinate ${\bf z}={\bf y}_1-{\bf y}_2$.
The functions ${\psi}_{nl}$ simultaneously
 with pionium energy levels ${\varepsilon}_{nl}$ are determined by the
 Schr\"odinger equation \cite{shl}
\begin{equation}
-\frac{1}{m}{\bigtriangledown}^{2}{\psi}_{nl}({\bf z}) + U({\bf
z}){\psi}_{nl}({\bf z}) = {\varepsilon}_{nl}{\psi}_{nl}({\bf z})
\label{4}
\end{equation}
with relevant boundary conditions at $z=0 , z\rightarrow\infty$. Here
$m=139.57 MeV$ is the ${\pi}^{\pm}$-meson mass \cite{mm} . We utilize the
units $c=h=1$.  For the pure-Coulomb point-like interaction
\begin{equation}
U(z) = - \frac{\alpha}{z} \label{6} \end{equation}
the ground state wave
function ${\psi}_{10}\equiv\psi$, properly normalized, and energy
${\varepsilon}_{10}\equiv\varepsilon$ are known \cite{shl} to be
\begin{equation}
{\psi}(z) = \frac{1}{\sqrt{4\pi}} \cdot \sqrt{\frac{a^3}{2}} e^{-za/2} , \; \;
{\varepsilon} = -\frac{m{\alpha}^2}{4} ,
\label{6a}
\end{equation}
where $a=m\alpha$ and $2/a$ is ``Bhor radius". Consequently, we denoute
 $|{\cal D}_{10}> \equiv |{\cal D}>$
Henceforth, we consider this
pionium ground state decay. The $\pi \pi$-interaction of the type
\cite{w1,w2,g1} including dependence on pion momenta being put to use in our
further calculations, the finite pion size $r_0$ emerges to come into the
picture, which we allow for in due course replacing (\ref{6}) by the
electrostatic potential between two homogeneously charged spheres, $z$ being
the distance between their centrepoints, that explicit expression, a bit
long, is set out in Ref.  \cite{cr}. Magnitude of the quantity $r_0$ itself
have been estimated in some theoretical and experimental investigations
\cite{rpa,rpw}, accordingly which we have adopted $r_0=0.6 fm$ as realistic.
If anything, it my be noted the calculations with generalized, but yet
instantaneous potential accounting for the relativistic corrections up to
$(1/c^2)$-order (the kind of Breit potential \cite{ll,ahb}) would not provide
the additional difficulties of principle.

In our present calculation, the $\pi \pi$-interaction inducing  the
${\pi}^{+}{\pi}^{-}\rightarrow 2{\pi}^{0}$ transition is specified by well
known Weinberg lagrangian
\begin{equation}
{\hat{\cal L}}_{\pi\pi}(x) = -\frac{1}{(2f_{\pi})^2}[\partial_{\mu}{\bfgr
\pi}(x)\partial^{\mu}{\bfgr \pi}(x) - {\beta}{\bar m}^{2}({\bfgr
\pi}(x))^2]{\bfgr \pi}^{2}(x)
\label{1}
\end{equation}
\setlength{\unitlength}{1cm}
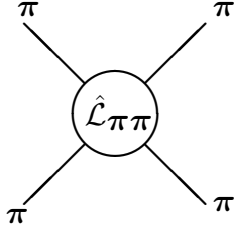
\begin{picture}(16.,3.)
\thicklines
\put(8,1.5){\circle{1.1}}
\put(7.6,1.35){$\hat {\bf\cal L}_{\bfgr \pi \bfgr \pi}$}
\put(8.405,1.905){\line(1,1){0.8}}
\put(8.405,1.095){\line(1,-1){0.8}}
\put(7.595,1.905){\line(-1,1){0.8}}
\put(7.595,1.095){\line(-1,-1){0.8}}
\put(9.3,2.8){$\bfgr \pi$}
\put(9.3,0.2){$\bfgr \pi$}
\put(6.7,2.8){$\bfgr \pi$}
\put(6.55,0.05){$\bfgr \pi$}
\end{picture}

elaborated and scrutinized in Refs. \cite{w1,w2,g1}. Here $f_{\pi}=92.4 MeV$
\cite{mm}. Dependence of the results of calculations on the parameters
${\beta} , \bar m$ in the chiral symmetry violating term will be discussed in
the last Section.

Let us recall the validity of the lagrangian (\ref{1}) have
been inferred from processing the experimental data on the
$N\pi\rightarrow N\pi\pi$ reaction, see Refs. \cite{bbo,bb2,bb3,bb1},
at least for not very high pions energies.

The difference of the masses of a charge pion, $m=139.57 MeV$, and
neutral one, $m_{0}=134.98 MeV, \; {\Delta}m=m-m_{0}=4.59 MeV$ being
greater than pionium binding energy $\varepsilon$, the initial
${\pi}^{+}{\pi}^{-}$ bound state ${|\cal D>}$ transition  into the
final two ${\pi}^{0}$ state turns out to be possible via processes
presenting by (\ref{5o}, \ref{1}). All the effective interactions between pion
(charge and neutral) and pionium fields are described by the total
interaction lagrangian
\begin{equation}
{\hat{\cal L}}_{tot} = {\hat{\cal L}}_{\cal D} + {\hat{\cal
L}}_{\pi\pi},
\label{l7}
\end{equation}
which determines eventually pionium lifetime $\tau$.

\begin{center}
{\large \bf 4. Pionium decay amplitude.}
\end{center}

The matrix element
\begin{equation}
{\cal S}_{{\pi}^{0}{\pi}^{0}{\cal D}} = <{\pi}^{0}{\pi}^{0}
|{\hat {\cal S}}|{\cal D}>
\label{8}
\end{equation}
of the ${\hat {\cal S}}$-matrix dictated by the lagrangian (\ref{l7})
determines the initial pionium state ${|\cal D>}$ decay into two final
${\pi}^{0}$. To the first order in ${\hat{\cal L}}_{\pi\pi}$ (\ref{1}),
the $\cal S$-matrix element (\ref{8}) takes the form (see, for
instance, \cite{ll,sw})
\begin{eqnarray}
{\cal S}_{{\pi}^{0}{\pi}^{0}{\cal D}}^{1} = -\int d{\bf R} \int d{\bf
z} \int dt \int d^{4}x <{\pi}^{0}{\pi}^{0}|{\hat T}[
{\hat{\cal L}}_{\cal D}({\bf R, z}, t) \cdot {\hat{\cal
L}}_{\pi\pi}(x)]|{\cal D}> = \nonumber \\
\nonumber \\
=\frac{i{\cal N}8}{(2f_{\pi})^{2} \, 2 \,
\sqrt{2E_{\lambda}{\varepsilon}_{1}{\varepsilon}_{2}}}
\int d{\bf R} \int d{\bf z} \int dt \int d^{4}x U({\bf
z}){\psi}_{\lambda}({\bf z})
{\{} 2 {\beta} {\bar m}^{2} -
({\varepsilon}_1 {\varepsilon}_2 -{\bf p}_1 {\bf p}_2)+{\partial}_{x\mu}
 {\partial}_{x}^{\mu} {\}}{\times} \nonumber \\
\nonumber \\
\times{\it D}({\bf R}+{\bf z}/2-{\bf x},t-x_0)\cdot {\it D}({\bf
R}-{\bf z}/2-{\bf x},t-x_0)\cdot e^{-iE_{\lambda}+i{\bf P R}} \cdot
e^{ix_{0}({\varepsilon}_1+{\varepsilon}_2)-i{\bf x}({\bf p}_{1}+{\bf
p}_{2})},\, \label{9}
\end{eqnarray}
\setlength{\unitlength}{1cm}
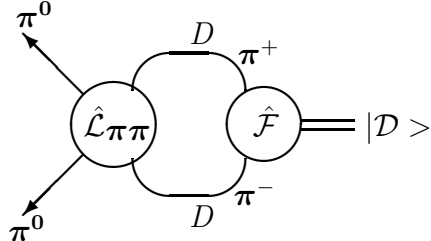
\begin{picture}(16.,3.)
\thicklines
\put(7,1.5){\circle{1.1}}
\put(6.6,1.35){$\hat {\bf\cal L}_{\bfgr \pi \bfgr \pi}$}
\put(9,1.5){\circle{1.}}
\put(8.85,1.35){$\hat {\bf\cal F}$}
\thinlines
\multiput(9.5,1.43)(0,0.12){2}{\line(1,0){0.7}}
\put(10.34,1.35){${\bf |\cal D>}$}
\thicklines
\put(5.7,2.8){${\bfgr \pi}^{\bf 0}$}
\put(5.6,-.05){${\bfgr \pi}^{\bf 0}$}
\put(6.595,1.905){\vector(-1,1){0.8}}
\put(6.595,1.095){\vector(-1,-1){0.8}}
\put(8,1.95){\oval(1.5,1.0)[t]}
\put(8,1.05){\oval(1.5,1.0)[b]}
\put(8,0.13){$\it D$}
\put(8,2.57){$\it D$}
\put(8.65,2.3){${\bfgr \pi}^{\bf +}$}
\put(8.6,0.43){${\bfgr \pi}^{\bf -}$}
\end{picture}

where ${\hat T}$ is usual time-ordering operator and
${\varepsilon}_{1,2} , {\bf p}_{1,2}$ denote final ${\pi}^{0}$ energies
and momenta. Certainly, when necessary, the high ${\hat{\cal L}}_{\pi
\pi}$-order contributions in (\ref{8}) could be allowed for in usual
way. These terms, if calculated, would render, in particular, the effect
of strong $\pi \pi$-interaction on the pionium state. In course of our
to-day calculation, we restricted ourselves by accounting for the first
 ${\cal L}_{\pi \pi}$-order. If anything, it may be recall the
 analysis of $N{\pi}{\rightarrow}N{\pi \pi}$ reaction was carried out in
 Refs. \cite{bbo,bb2,bb3,bb1}, as a matter of fact, in the same first
 order in  ${\cal L}_{\pi \pi}$ approximation. For the ground state
 pionium decay at rest, the relations hold
\begin{eqnarray}
{\bf P}=0, \; \; E_{n{\bf P}}=E=2m+\varepsilon , \; \; {\varepsilon}_1=
{\varepsilon}_2={\varepsilon}_{0}=E/2, \nonumber \\
{\bf p}_1=-{\bf p}_2, \; \;
 |{\bf p}_1|=|{\bf p}_0|=p_{0}=\sqrt{(E/2)^2-m^2_0} \; ,
\label{po}
\end{eqnarray}
and the Eq. (\ref{9}) reduces as follows
\begin{eqnarray}
{\cal S}_{{\pi}^{0}{\pi}^{0}{\cal D}}^{1} = i(2\pi)^4 \cdot
{\cal T}_{{\pi}^{0}{\pi}^{0}{\cal D}} \cdot \delta ({\bf p}_1+{\bf
p}_2) \delta ({\varepsilon}_2+{\varepsilon}_1-E) \, , \nonumber \\
  \label{10} \\
{\cal T}_{{\pi}^{0}{\pi}^{0}{\cal D}} = \frac{1}{(2\pi)^4}\frac{8N}
{(2f_{\pi})^2E\sqrt{2E}} \int d{\bf z}{\cdot}U({\bf z}){\cdot}{\psi}({\bf z})
\times \nonumber \\
\nonumber \\
\times\int d^4q\frac{-2\beta{\bar m}^2+m_0^2-E^2/2+q^2_0-{\bf q}^2-q_0E}
{[q_0^2-{\bf q}^2-m^2+i0]\cdot[(E-q_0)^2-{\bf q}^2-m^2+i0]}\cdot
e^{-i{\bf zq}}. \nonumber
\end{eqnarray}
Let us take cognizance of the quantities ${\bf q}^2, q_0^2$ emergence in
the nominator in (\ref{10}) which is due to the term
$$({\partial}_{\mu}
{\bfgr \pi}\cdot{\partial}^{\mu}{\bfgr \pi})({\bfgr \pi})^2$$
in the $\pi
\pi$-interaction (\ref{1}), this fact substantially affected the
integrand behaviour in (\ref{10}), especially at extremely  large $q$
values. Integrating over $dq_0$ and over directions of the vectors
${\bf q}$ and ${\bf z}$ having been carried out, the Eq. (\ref{10})
reduces to
\begin{eqnarray}
{\cal T}_{{\pi}^{0}{\pi}^{0}{\cal D}} = \frac{-i8{\cal N}}
{\pi (2f_{\pi})^2 E\sqrt{2E}} \int_{0}^{\infty} \frac{q \, dq}{{\omega}(q)}
 \cdot {\Huge [} 1-\frac{b}{q^2+c^2} {\Huge ]} \cdot
 \int_{0}^{\infty} dz{\cdot}U(z){\cdot}z{\cdot}{\psi}(z){\cdot}sin(qz),
\label{11}
\end{eqnarray}
where the notations are introduced:
$${\omega}(q)=\sqrt{q^2+m^2} , \; \;
c^2=m^2-(E/2)^2 , \; \; b=(-2{\beta}{\bar m}^2+m_0^2+m^2-E^2)/2$$
It is not difficult ot realize the behaviour of the integrand in (\ref{11})
at large momenta, $q\rightarrow\infty$, and subsequently the convergence of
the integral in (\ref{11}) itself are governed by the behaviour of the
quantity $z U(z) {\psi}(z) sin(qz)$ when $z$ value tends to zero,
$z\rightarrow 0$.  There is to calculate the contributions arising from two
terms in brackets in integrand (\ref{11}): from ``unit", $1$, and from
$b/(q^2+c^2)$. We take up firstly integrating the term with ``unit" and then
set out the integral with quantity $b/(q^2+c^2)$.

Not hard thing is to become convinced the integral in
(\ref{11}) with ``unit" in brackets would diverge logariphmically, if the pure
point-like Coulomb values (\ref{6}), (\ref{6a}) were adopted for quantities
$U(z), {\psi}(z)$ in (\ref{11}). This divergency emerges because a pion
size is neglected. To remove this puzzling, but spurious contradiction we
allow for the finite pion size $r_0, \;  r_0 a\ll r_0 m\ll 1$ in course of
calculating this integral, $U(z)$ being the electrostatic potential between
two homogeneously charged spheres of the radius $r_0$ \cite{cr}, as discussed
already after Eq. (\ref{6a}). Then, integrating over $dq$ having been
performed, the integral in (\ref{11}) originating due to the ``unit" in
brackets transforms to (see Ref. \cite{rg})
\begin{eqnarray}
-\int_{0}^{\infty} dz\cdot z U(z) {\psi}(z) \frac{d}{dz}K_0(mz) =
 - z U(z) {\psi}(z) K_0(mz){\Large |^{{\infty}}_{{0}}} + \nonumber \\
\int^{2r_{0}}_0 dz K_0(mz) {\psi}(z) \frac{d}{dz}[z U(z)] + \nonumber \\
 {\alpha}\int_{2r_0}^{\infty} dz K_0(mz)\frac{d}{dz}{\psi}(z) +
 \int^{2r_0}_0 dz K_0(mz) z U(z)\frac{d}{dz}{\psi}(z),
\label{12}
\end{eqnarray}
where $K_0(z)$ is the Mackdonald's cylindrical function (see \cite{rg}). The
first term in the righthand side (\ref{12}) vanishes due to ${\psi}(z)\sim
e^{-az/2}\rightarrow 0$ when $z\rightarrow\infty$, and it disappears at $z=0$
owing to $zU(z)=0$ at $z=0$ because, in turn, the potential $U(z)$ for finite
size charged particles has got at $z=0$ a finite value $U(0)$, in
particular, for the afore adopted potential of homogeneously charged spheres
$U(0)=-6\alpha /(5r_0)$. Further, in our treatment, we are on the point to
carry out all the calculations in the lowest $\alpha$-order. All the
expression (\ref{11}) (as well as (\ref{12})) is proportional to $\alpha
\sqrt{{\alpha}^3}$ due to the $\alpha$-dependence of the functions $U, \psi$.
Calculating the integrals  in (\ref{11}), (\ref{12}), we retain only the
terms which besides this $\alpha$-dependence are inversely proportional to
$\alpha , \; \sim 1/\alpha ,$ and $\alpha$-independent. Even so, we retain in
the asymptotic expansion in $r_0$ only the terms $\sim ln(r_0)$, but drop out
the terms $\sim r_0^n, n\geq 1$.  Consequently, the second and third
integrals in righthand side in (\ref{12}) are realized to be neglected.
Indeed, at $z\geq 2r_0$ in the second integral, the function $\psi (z)$
behaves likely (\ref{6a}), $\sim e^{-za'/2}$, the quantity $a'$ being of the
same order in $\alpha$ as $a, a'\approx a=m\alpha$ (see, for instance, Refs.
\cite{cr,pus}). Then, we have got $d{\psi}(z)/dz \sim {\alpha}m{\psi}(z)/2$,
and, subsequently, this integral gets the additional factor of $\alpha$ and
must be dropped out.  Then, in the third integral in righthand side
(\ref{12}), for $z\leq 2r_0\ll 2/a$, the function ${\psi}(z)$ varying
smoothly (see, for instance, Refs.  \cite{pus,cr}), ${\psi}(z)\sim{\psi}(0)
 (1+z a'')$, where $a''$ is of the same order as $a$, the derivative
 $d{\psi}(z)/dz\approx{\psi}(0){\cdot}a$, so as the whole integral comes out
 $\sim\alpha r_0 a {\psi}(0)$ and is to be omitted as well. Thus, eventually,
 there is to calculate the first integral in the righthand side of
 (\ref{12}). Its upper limit turned out to be $2r_0$ because
 $\frac{d}{dz}[U(z) z]=0$ at $z\geq 2r_0, \, U(z)$ being point-like Coulomb
 potential $-{\alpha}/z$ when $z\geq 2r_0$. For this $z$ values, the
 relations $r_{0}a\ll 1, \; r_{0}m\ll 1$ being valid, the replacements hold
 true \begin{equation} {\psi}(z) = {\psi}(0), \; \; K_{0}(mz) = - ln(mz/2) -
C \label{13} \end{equation} with the accuracy up to order $\sim r_{0}a, \;
\sim r_{0}m$. Here $C\approx 0.577$ is Euler constant (see \cite{rg}). Then,
the approximation (\ref{13}) being put to use, the first integral in
righthand side in (\ref{12}) is straightforward calculated, and the whole
expression (\ref{12}) results in \begin{equation} {\psi}(0) [ \alpha
 (ln(mr_{0}) + C) + \bar U ], \; \; \; \bar U = \int_{0}^{2r_{0}} dz U(z)
 \label{14} \end{equation} The quantity $\bar U$ is calculated accordingly to
Ref. \cite{cr} which gives ${\bar U}\approx -{\alpha}{\cdot}(3/2)$.

While treating the integral with the term $b/(c^2+q^2)$ within brackets in
(\ref{11}), the presence of an additional $q^2$ in the denominator provides
this integral convergence even without the finite pion size $r_0$ being taken
into account. This does mean to say that asymptotic expanding this integral
in $r_0$ begins with a term $\sim r_0$ which is beyond our to-day accuracy, as
has been presumed abouv. Then, the Eqs. (\ref{6}, \ref{6a}) being adopted,
this integral in the lowest $\alpha$-order reduces as follows
\begin{equation}
\sqrt{\frac{a^3}{8\pi}} b \alpha \int_{0}^{\infty} \frac{dq\cdot q^2}
{(q^2+(a/2)^2) (q^2+c^2) {\omega}(q)} = \sqrt{\frac{a^3}{8\pi}} \frac{b}{m}
[\frac{\pi}{2}-\alpha ]
\label{15}
\end{equation}
After all, with allowance for results (\ref{14}, \ref{15}), the transition
amplitude (\ref{11}) takes the form
\begin{equation}
{\cal T}_{{\pi}^{0}{\pi}^{0}{\cal D}} = -\frac{i8{\cal N}}
{{\pi}(2f_{\pi})^{2}E \sqrt{2E}} \cdot \sqrt{\frac{a^3}{8\pi}} \cdot
 {\Huge [} \frac{b}{m^2} (\frac{\pi}{2} - \alpha ) + \alpha {\Huge (}
ln(mr_{0}) + C ) + \bar U {\Huge ]} \label{16} \end{equation}

The normalization factor $\cal N$ residing in the Eqs. (\ref{5}, \ref{9} -
\ref{11}, \ref{16}) is to determined by equating the energy $E$ of the state
$|{\cal D}>$ of a pionium at rest and the expectation value in the
 $|{\cal D}>$-state of the operator of the ${\hat {\cal T}}^{00}$-component
 of energy-momentum tenser of a charge (complex) pion field:
\begin{eqnarray}
E = <{\cal D}| \; {\hat T} ({\hat {\cal T}}^{00} {\hat {\cal S}}_{\cal D})
|{\cal D}> \label{17} , \\
{\hat {\cal T}}^{00}({\xi}_{0},{\bfgr \xi}) =
-{\large [}\frac{{\partial}{\pi}^{-}({\xi})}{{\partial}{\xi}_{0}}\cdot
\frac{{\partial}{\pi}^{+}({\xi})}{{\partial}{\xi}_{0}} +
\frac{{\partial}{\pi}^{-}({\xi})}{{\partial}{\bfgr \xi}}\cdot
\frac{{\partial}{\pi}^{+}({\xi})}{{\partial}{\bfgr \xi}} +
m^2 {\pi}^{-}(\xi ){\pi}^{+}(\xi ){\large ]} ,
\label{18}
\end{eqnarray}
where the ${\hat {\cal S}}_{\cal D}$-matrix is dictated by the lagrangian
(\ref{5}), so as
\begin{eqnarray}
E = -\frac{1}{2}<{\cal D}| \; {\hat T} {\Large [} {\hat {\cal
T}}^{00}({\xi}_{0},{\bfgr \xi}) \int d{\bf y}_{1} d{\bf y}_{2} dt {\hat {\cal
L}}_{\cal D}({\bf y}_{1},{\bf y}_{2},t) \int d{\bf y}^{,}_{1} d{\bf
 y}^{,}_{2} dt^{,} {\hat{\cal L}}_{\cal D}({\bf y}^{,}_{1},{\bf
y}^{,}_{2},t^{,}) {\Large ]} |{\cal D}> , \label{18a}
\end{eqnarray}
and, for
clarity's sake, the expression is worth of being displayed by the usual
diagram
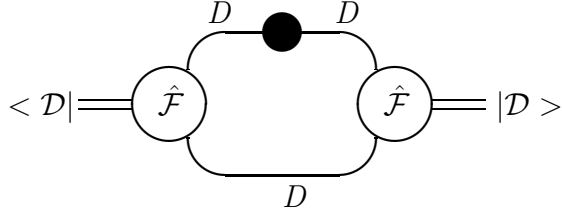
\setlength{\unitlength}{1cm}
\begin{picture}(16.,3.)
\thicklines
\put(9.5,1.5){\circle{1.}}
\put(9.35,1.35){$\hat {\bf\cal F}$}
\thinlines
\multiput(10,1.43)(0,0.12){2}{\line(1,0){0.7}}
\put(10.84,1.35){${\bf |\cal D>}$}
\thicklines
\put(6.5,1.5){\circle{1.}}
\put(6.35,1.35){$\hat {\bf\cal F}$}
\thinlines
\multiput(5.3,1.43)(0,0.12){2}{\line(1,0){0.7}}
\put(4.35,1.35){${\bf <\cal D |}$}
\thicklines
\put(8,1.95){\oval(2.5,1.0)[t]}
\put(8,1.05){\oval(2.5,1.0)[b]}
\put(8,2.45){\circle*{0.6}}
\put(8,0.13){$\it D$}
\put(7,2.57){$\it D$}
\put(8.7,2.57){$\it D$}
\end{picture}

where the solid disk stands for the ${\hat {\cal T}}^{00}$ operator. The
values (\ref{6},\ref{6a}) asserted for the point-like pion being adopted, the
straightforward calculation of (\ref{18a}) results in
\begin{equation}
E = \frac{2{\alpha}^{2} {\cal N}^{2} a^3}{E\pi} \int_{0}^{\infty}
\frac{dq{\cdot}q^2}{{\omega}(q)} \frac{E^2 + 4{\omega}^{2}(q)}{(q^2 + (a/2)^2
)^2\cdot (4{\omega}^{2}(q) - E^2 )}
\label{21}
\end{equation}
which apparently shows up no divergency by integrating over $dq$ which is due
to the integrand steep enough decrease at $q\rightarrow\infty$ on the account,
in turn, of the high power of $q$ in the denominator of (\ref{21}). Evaluating
(\ref{21}), we are to retain only the terms of the lowest $\alpha$-order: the
$\alpha$-independent terms and terms $\sim\alpha$ (if they would have
appeared), omitting the terms $\sim\alpha ^{n}, n>1$. Then, Eq.
(\ref{21}) reduces to
\begin{equation}
E = \frac{{\cal N}^2}{E m} , \; \; \; {\cal N}^2 = 4 m^3
\label{22}
\end{equation}
If anything, for verification's sake, the ${\cal N}$ value can be obtained by
equating the expectation value of the particles number operator related to
zeroth component of charged pion field current)
 $$ {\hat N} = \sum_{{\bf p}}
 [ {a^{+}}_{{\bf p}} a_{{\bf p}} + b^{+}_{{\bf p}} b_{{\bf p}} ] $$
 (see Eqs.
(\ref{1ab})) in the pionium state $|{\cal D}>$ and the pions number $N=2$,
that is from the equation
$$ <{\cal D}| {\hat T} ({\hat N} {\hat {\cal
S}}_{\cal D}) |{\cal D}> = 2 . $$
All the calculations having been carried
out in due course, we arrive  at the same ${\cal N}$ value (\ref{22}). Let us
take cognizance of the fact that the righthand side in Eqs. (\ref{18} -
\ref{21}) proves having got no terms $\sim\alpha$, its expansion in $\alpha$
starting with a term $\sim\alpha ^2$.  Evidently, it must be just so because
the quantity $E=2m-m{\alpha}^2/4$ in the lefthand side does not include terms
$\sim\alpha$.

\begin{center}
{\large \bf 5. Pionium life-time calculation results and concluding remarks.}
\end{center}

Thus, we have at our disposal the expression (\ref{16}) for the transition
amplitude with $\cal N$ defined by Eq. (\ref{22}). Then, we acquire in the
usual way (see, for instance, Ref. \cite{ll}) the total probability $W$ of
 a pionium conversion into two ${\pi}^0$, that is the inverse pionium lifetime
$\tau$
\begin{equation} W = \frac{1}{\tau} = \frac{a^3 {\cdot} p_0 {\cdot}  m^3
 {\cdot}  {\tilde b}^2
}{( 2 f_{\pi} )^4 \, 2{\pi}^2 E^2 } [1 - \frac{4\alpha}{\pi} (1 - \frac{{\bar
U}/{\alpha} + ln(mr_0) + C}{\tilde b}) ], \label{23}
\end{equation}
where ${\tilde
b}=[-2{\beta}{\cdot}({\bar m}/m)^2-(m^0/m)^2-3]/2$ and all the other
quantities have been set forth above.

Let us now inquire into how the $\tau$
value (\ref{23}) depends on the values of ${\bar m} , \beta$ which reside in
the chiral symmetry violating term in lagrangian (\ref{1}). Let firstly
${\bar m}=m_0$, then we gain for the $\beta$ values ${\beta}=1/2 , \;
{\beta}=1/3 , \; {\beta}=1/4$ asserted in Refs. \cite{w1,w2,g1}
$${\tau}_{m_{0},1/2} = 4.95{\cdot}10^{-15} sec , \; \; {\tau}_{m_{0},1/3} =
6.18{\cdot}10^{-15} sec , \; \; {\tau}_{m_{0},1/4} = 6.90{\cdot}10^{-15} sec
$$
Thus, the dependence
of $\tau$ on $\beta$ is thought to be sizeable, the deviations of these
$\tau$ values from each other amounting $\approx 15 \%$.  On the other hand,
if we adopt ${\bar m}=m$ instead ${\bar m}=m_0$, we shall have got
$${\tau}_{m,1/2} = 4.71{\cdot}10^{-15} sec$$
which deviates from ${\tau}_{m_{0},1/2}$ by about $5 \%$. Let us also note
the second term within brackets in (\ref{23}) amounts $\approx 2 \%$ to the
whole $W$ value.

Our result proves do not contradict the up to now estimation $\tau =
2.9_{-2.1}^{+\infty}{\cdot}10^{-15}sec$ set out in Ref. \cite{n3}. It might be
instructive to recall the results of $\tau$ calculation obtained in the
previous investigations, surveyed in Section 2, appear to be some smaller as
compare to our one. For instance, the value ${\tau}=2.72{\cdot}10^{-15}sec$
has been asserted in Ref.  \cite{wy} and ${\tau}=3.2{\cdot}10^{-15}sec$ in
Ref.  \cite{eul}.

The investigation carried out makes us realize the pionium life-time (as well
as its other properties) does depend crucially on the form of the genuine
$\pi\pi$-interaction, but not simply just on the free pions scattering
 lenghts only.  Thus, the pionium decay as being due to the most plausible
concise Weinberg lagrangian (\ref{1}) having been studied, the investigations
pursuing  other present-day trustworthy $\pi\pi$-interaction descriptions are
very desirable and instructive. If the consistent $\tau$ calculation in
framework of a certain method of $\pi\pi$-interaction description (see, for
instance Refs.  \cite{lg,nj,bb4}) is carried out and, subsequently, its
result is confronted to the experimental $\tau$ value, the validity of this
method will come to light. We are on the point of inquiring into the various
$\pi\pi$-interaction representations in the course of pionium lifetime
studying.


\newpage
\begin{thebibliography}{99}
\bibitem{w1} S. Weinberg, Phys. Rev. Lett. 18 (1967) 188.\\
S. Weinberg, Physica A 96 (1979) 327.
\bibitem{w2} S. Weinberg, Phys. Rev. Lett. 17 (1966) 616.
\bibitem{g1} P. Chang and F. G\"ursey, Phys. Rev. 164 (1967) 1752.\\
J. Schwinger, Phys. Lett. B 24 (1967) 473.\\
W. A. Bardin, L. Brown, B. W. Lee and H. T. Neih, Phys. Rev. Lett. 18 (1967)
1170.
\bibitem{lg} J. Gasser and H. Leutwyler, Ann. of Phys. 158 (1984) 142.\\
U.-G. Meissner, Rep. Prog. Phys. 56 (1993) 903.\\
G. J. Stephenson, K. Maltman and T. Goldman, Phys. Rev. D 43 (1991) 860.\\
J. Stern, H. Sazdjian and N. H. Fuchs, Phys. Rev. D 47 (1993) 3814.
\bibitem{nj} A. A. Andrianov, V. A. Andrianov, Yu. V. Novozhilov and V. Yu.
Novozhilov, Phys. Lett. B 186 (1987) 401.
\bibitem{c1} A. Iyonov et al. ((EMU05) Collaboration), Nucl. Phys. A 544
(1991) 55c.\\
R. Rapp, J. Wambach, Phys. Rev. C 53 (1996) 3057.\\
G. G. Bunatian, J. Wambach, Phys. Lett. B 336 (1994) 290.\\
G. G. Bunatian, Sov. J. Nucl. Phys. 31 (1980) 631; 41 (1985) 560.
\bibitem{c2} G. G. Bunatian and I. N. Mishustin, Nucl. Phys. A 404 (1983)
525.
\bibitem{bbo} Yu. A. Batusov, S. A. Buniatov, V. M. Sidorov and V. A. Yarba,
 Sov. J. Nucl. Phys. 1 (1965) 526.
\bibitem{bb2} G. Ecker and J. Honerkampf, Nucl. Phys. B 62 (1973) 509; Nucl.
 Phys. B 52 (1973) 211.\\
H. Lehmann and H. Trute, Nucl. Phys. B 52 (1973) 280.\\
T. H. Truong, Phys. Lett. B 99 (1981) 154.\\
O. J\"ackel, H.-W. Ortner and M. Dilling, Nucl. Phys. A 511
(1990) 733; Nucl. Phys. A 541 (1992) 675.
\bibitem{bb3} B. R. Martin, D. Morgan and G.Shaw, Pion-Pion Interaction in
Particle Physiks (Academic Press, New York, 1977).
\bibitem{bb1} A. A. Bel'kov and S. A. Buniatov, Sov. J. Particles and Nuclei
13 (1982) 5.
\bibitem{bb4} D. Lohse, J. W. Durso, K. Holinde and J. Speth, Nucl. Phys. A
516 (1990) 513.
\bibitem{kls} L. Rosselet, Phys. Rev. D 15 (1977) 574.
\bibitem{com} E. D. Commins and P. H. Bucksbaum, Weak Interactions of Leptons
and Quarks (Cam. Univ. Press, Cam., England, 1983).
\bibitem{n1} J. L. Uretsky and T. R. Palfrey, Phys. Rev. 121 (1961) 1798.\\
S. M. Bilenky, Hyuen Van Heu, L. L. Nemenov and F. G. Tkebuchava, Sov. J.
Nucl. Phys. 10 (1969) 812.
\bibitem{n2} L. L. Nemenov, Sov. J. Nucl. Phys. 15 (1972) 1047; 41 (1985)
980.
\bibitem{n3} L. G. Afanasyev et al., Phys. Lett. B 308 (1993) 200; B 338
(1994) 478.
\bibitem{dg} S. Deser, M. I. Goldberger, K. Baumann and W. Thirring, Phys.
Rev. 96 (1954) 774.\\
H. Bethe and F. Hoffman, Mesons and Fields, V. 2 (Row, Peterson and Co.,
Evanston, Illinois, 1955) p. 103.
\bibitem{dg1} N. Byers, Phys. Rev. 107 (1957) 843.
\bibitem{dg2} T. L. Trueman, Nucl. Phys. 26 (1961) 57.
\bibitem{wy} S. Wycech and A. M. Green, Nucl. Phys. A 562 (1993) 446.
\bibitem{ras} U. Moor, G. Rasche and W. S. Woolcock, Nucl. Phys. A 587
(1995) 747.
\bibitem{eul} G. V. Efimov, M. A. Ivanov and V. E. Lubovitsfij, Sov. J.
Nucl. Phys. 44 (1986) 296.
\bibitem{bpt} A. A. Bel'kov, V. N. Pervushin and F. G. Tkebuchava, Sov. J.
Nucl. Phys. 44 (1986) 300.
\bibitem{lub} V. E. Lubovitskij and A. G. Rusetsky, Phys. Lett. B 389 (1996)
181.
\bibitem{sil} Z. Silagadze, JETPh Lett. 60 (1994) 689.
\bibitem{ll} E. M. Lifshiz, L. P. Pitaevsky, Relativistic Quantum Theory,
Parts 1 and 2 (Nauka, Moscow, 1971).
\bibitem{sw} S. Shweber, An Introduction to Relativistic Quantum Field Theory
(Row, Peterson and Co., N. Y., 1961).
\bibitem{ab}
A. B. Migdal, FInite Fermi System Theory and Atomic Nuclei Properties (Nauka,
Moscow, 1965).
\bibitem{shl} L. D. Landau and E. M. Lifshiz, Quantum Mechanics (FM, Moscow,
1963).\\
L. I. Schiff, Quantum Mechanics (MeCraw-Hill Book Company, England, 1955).
\bibitem{mm} Rev. Part. Prop., Phys. Rev. D 50, Patr I (1994) 1177.
\bibitem{cr} G. C. Oades and G. Rasche, Nucl. Phys. B 20 (1970) 333.
\bibitem{rpa} S. R. Amendolia, Phys. Lett. B 244 (1984) 469; Nucl. Phys. B
138 (1986) 168.
\bibitem{rpw} V. Bernard, R. Brockman and W. Weise, Nucl. Phys. A 434 (1985)
685.
\bibitem{ahb} A. I. Achiezer and V. B. Berestezky, Quantum Electrodynamic
(FM, Moscow, 1959).
\bibitem{rg} I. S. Gradstein and I. M. Rizik, Tables of Integrals, Sums,
Serieses and Products (FM, Moscow, 1963).
\bibitem{pus} I. P. Pustovalov, ZETP 36 (1959) 1806.
\end{thebibliography}
\end{document}